\documentclass[aps, prd, floatfix, nofootinbib, showpacs, twocolumn, superscriptaddress]{revtex4-1}

\usepackage{amsmath,amsfonts,amssymb,bm}
\usepackage{graphicx}
\usepackage{color}
\usepackage{subfigure}
\usepackage{multirow}
\usepackage{textcomp}
\usepackage{slashed}

\usepackage[mathscr,scaled=1.15]{urwchancal}
\DeclareFontFamily{OT1}{pzc}{}
\DeclareFontShape{OT1}{pzc}{m}{it}%
{<-> s * [1.15] pzcmi7t}{}
\DeclareMathAlphabet{\mathpzc}{OT1}{pzc}{m}{it}

\definecolor{purple}{rgb}{0.5,0,0.5}
\definecolor{blue}{rgb}{0.0,0,0.9}
\definecolor{prdblue}{rgb}{0.133,0.118,0.498}
\usepackage[colorlinks=true, pdfstartview=FitV, linkcolor=prdblue, citecolor= prdblue, urlcolor=prdblue]{hyperref}

\begin{document}

\title{
Pseudoscalar glueball mass: a window on three-gluon interactions}

\author{E.~V.~Souza}
\email{emanuel.veras@ifpi.edu.br}
\affiliation{Federal Institute of Education, Science and Technology of Piau{\'{\i}}, 64605-500, Picos, Piau{\'{\i}}, Brazil}
\affiliation{University of Campinas - UNICAMP, Institute of Physics ``Gleb Wataghin'',
13083-859 Campinas, S\~ao Paulo, Brazil}

\author{M.~N.~Ferreira}
\email{mnarciso@ifi.unicamp.br}
\affiliation{University of Campinas - UNICAMP, Institute of Physics ``Gleb Wataghin'',
13083-859 Campinas, S\~ao Paulo, Brazil}
\affiliation{Department of Theoretical Physics and IFIC, University of Valencia and CSIC, E-46100, Valencia, Spain}

\author{A.~C.~Aguilar}
\email[]{aguilar@ifi.unicamp.br}
\affiliation{University of Campinas - UNICAMP, Institute of Physics ``Gleb Wataghin'',
13083-859 Campinas, S\~ao Paulo, Brazil}

\author{J.~Papavassiliou}\email[]{Joannis.Papavassiliou@uv.es}
\affiliation{Department of Theoretical Physics and IFIC, University of Valencia and CSIC, E-46100, Valencia, Spain}

\author{C.~D.~Roberts}
\email[]{cdroberts@nju.edu.cn}
\affiliation{School of Physics, Nanjing University, Nanjing, Jiangsu 210093, China}
\affiliation{Institute for Nonperturbative Physics, Nanjing University, Nanjing, Jiangsu 210093, China}

\author{S.-S.~Xu}
\email[]{xuss@njupt.edu.cn}
\affiliation{College of Science, Nanjing University of Posts and Telecommunications, Nanjing 210023, China}

\date{06 December 2019}

\begin{abstract}
\hspace*{-\parindent}\mbox{\sf Abstract}.
In pure-glue QCD, gluon-gluon scattering in the $J^{PC}=0^{-+}$ channel is described by a very simple equation, especially if one considers just the leading contribution to the scattering kernel.  Of all components in this kernel, only the three-gluon vertex, $V_{\mu\nu\rho}$, is poorly constrained by contemporary analyses; hence, calculations of $0^{-+}$ glueball properties serve as a clear window onto the character and form of $V_{\mu\nu\rho}$.  This is important given that many modern calculations of $V_{\mu\nu\rho}$ predict the appearance of an infrared suppression in the scalar function which comes to modulate the bare vertex after the nonperturbative resummation of interactions.  Such behaviour is a peculiar prediction; but we find that the suppression is essential if one is to achieve agreement with lattice-QCD predictions for the $0^{-+}$ glueball mass.  Hence, it is likely that this novel feature of $V_{\mu\nu\rho}$ is real and has observable implications for the spectrum, decays and interactions of all QCD bound-states.
\end{abstract}

\maketitle

\noindent{\sf \textbf{1$\;$Introduction}}.
The Clay Mathematics Institute has established a ``Millennium Problem'' prize for proving that quantum $SU_c(3)$ gauge field theory is mathematically well-defined \cite{Jaffe:Clay}.  Supposing it is, then one corollary must be the dynamical generation of a mass gap, $\Delta$, in pure-gauge QCD and the attendant appearance of glueball bound-states.  There is strong numerical evidence in support of these outcomes, found especially in the \mbox{fact} that numerical simulations of lattice-regularised QCD (lQCD) predict $\Delta \gtrsim 1.5\,$GeV \cite{McNeile:2008sr} and a rich spectrum of glueball bound-states \cite{Chen:2005mg}.

\begin{figure}[t]
 	\includegraphics[clip, width=0.48\textwidth]{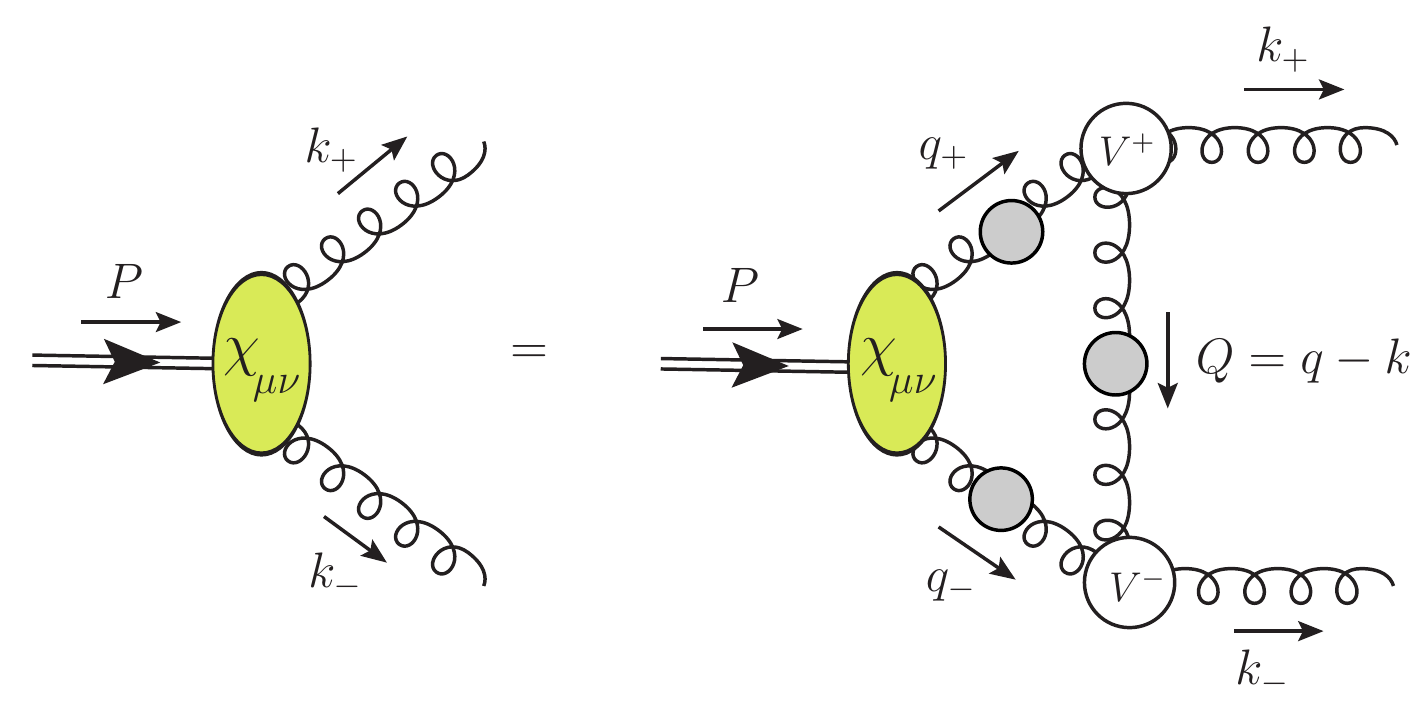}
\caption{\label{fig:BSE}
Leading contribution to the Bethe-Salpeter equation for a $0^{-+}$ glueball with total momentum $P$ in pure-glue QCD: ``springs'' -- gluon propagators, $D_{\mu\nu}(\ell)$; and ``open circles'' -- three-gluon vertices, $V_{\mu\nu\rho}$.
(${\mathpzc l}_\pm = \pm {\mathpzc l} +P/2$, ${\mathpzc l} = k,q$).}
\end{figure}

With lQCD having established benchmarks, continuum bound-state methods are being employed in attempts to develop an intuitive understanding of glueball emergence and structure, \emph{e.g}.\ Refs.\,\cite{Dudal:2010cd, Meyers:2012ka, Sanchis-Alepuz:2015hma, Meyer:2015eta}.
In pure-gauge QCD, the $J^{PC}=0^{-+}$ glueball is the simplest case because the Bethe-Salpeter equation has only one dynamical kernel, \emph{viz}.\ that describing gluon-gluon scattering, the leading contribution to which is illustrated in Fig.\,\ref{fig:BSE};
and the solution amplitude involves just one scalar function:
\begin{equation}
\chi_{\mu \nu}(k_+,k_-) = \epsilon^{\mu \nu \alpha \beta}k_{\alpha}P_{\beta}\mathcal{F}(k;P) \,.
\label{glupseudo}
\end{equation}
The $0^{-+}$ channel therefore presents the best opportunities for developing insights that are qualitatively insensitive to model details; and this is our goal herein.

\smallskip

\noindent{\sf \textbf{2$\;$Bethe-Salpeter Equation for \mbox{$ \mathbf 0^{\mathbf -+}$ Glueball}}}.
It is worth beginning with the observation that
Owing to Eq.\,\eqref{glupseudo}:
\begin{equation}
\label{BSTransverse}
k_{\pm \mu} \chi_{\mu \nu}(k_+,k_-) = 0 = k_{\pm \nu} \chi_{\mu \nu}(k_+,k_-) .
\end{equation}
Consequently, longitudinal gluon modes do not contribute to the bound-state; hence one can express the Bethe-Salpeter kernel quite generally using Landau gauge, with
\begin{equation}
D_{\mu\nu}({\mathpzc l}) = [\delta_{\mu\nu} - {\mathpzc l}_\mu {\mathpzc l}_\nu / {\mathpzc l}^2] \, D({\mathpzc l}^2)  =:  t_{\mu\nu}({\mathpzc l}) D({\mathpzc l}^2)\,.
\end{equation}
The merits of Landau gauge are widely known, \emph{e.g}.:
it is a fixed point of the renormalisation group;
a covariant gauge, which is readily implemented in simulations of lattice-regularised QCD;
and that gauge for which sensitivity to model-dependent differences between \emph{Ans\"atze} for the vertices is minimised.
Importantly, gauge covariance of Schwinger functions makes this last feature very valuable.  It means that if  \emph{Ans\"atze} are employed, then one need only formulate and compare them in Landau gauge.  Expressions in any other gauge contain no additional information: they are obtained using appropriate transformations which, when implemented correctly, must leave observables unaffected.

Shifting attention to the Bethe-Salpeter kernel, Fig.\,\ref{fig:BSE} shows that the leading contribution is defined by contractions of gluon propagators with three-gluon vertices and overall multiplication by the strong-coupling-squared, $g^2$.  (Notably, using Slavnov-Taylor identities, this combination is readily shown to be renormalisation-group invariant; hence, so is the solution.)  Combining the elements as indicated diagrammatically, the complete result is
\begin{align}
{\mathpzc K} = P_\delta P_\xi q_\gamma k_\zeta
\varepsilon_{\mu\nu\zeta\xi}
\varepsilon_{\alpha\beta\gamma\delta}
V_{\mu\alpha\rho}^+
V_{\nu\beta\sigma}^-
t_{\rho\sigma}(Q)\,,
\label{KKernel}
\end{align}
where $g^2$ and the scalar functions characterising $D_{\mu\nu}(\ell)$ have been omitted and, \emph{e.g}.\
$V_{\mu\alpha\rho}^\pm = V_{\mu\alpha\rho}(-k_\pm,q_\pm,\mp Q)$.

The three-gluon vertex may be expressed as follows:
\begin{equation}
V_{\alpha\mu\rho}(q,r,p) = V_{\alpha\mu\rho}^{\rm L}(q,r,p) + V_{\alpha\mu\rho}^{\rm T}(q,r,p)\,,
\end{equation}
where $V_{\alpha\mu\rho}^{\rm L}$ is that part which saturates the Slavnov-Taylor identities and $V_{\alpha\mu\rho}^{\rm T}$ is the completely transverse remainder.
The longitudinal term involves ten independent tensor structures:
\begin{equation}
V_{\alpha\mu\rho}^{\rm L}(q,r,p) = \sum_{i=1}^{10} X_i(q,r,p)\,\ell^i_{\alpha\mu\rho}\,,
\end{equation}
where $\{\ell^i_{\alpha\mu\rho}, i=1,\ldots,10\}$ are the momentum-dependent tensors defined in Ref.\,\cite{Aguilar:2019jsj}, Eq.\,(3.4).  Absent interactions, $X_1=X_4=X_7=1$ and all other scalar functions are zero; consequently,
\begin{align}
V_{\alpha\mu\rho}^{\rm L}&(q,r,p)  \to V_{\alpha\mu\rho}^{0\,{\rm L}}(q,r,p)\nonumber \\
&=  (q-r)_{\rho}\delta_{\alpha\mu} + (r-p)_{\alpha}\delta_{\mu\rho} + (p-q)_{\mu}\delta_{\alpha\rho}\,,
\label{Bare3Gluon}
\end{align}
which is the bare vertex.  In the presence of \mbox{interactions}, a realistic assessment of the probable nonperturbative forms of the functions $\{X_i\}$ in pure-gauge QCD is also provided in Ref.\,\cite{Aguilar:2019jsj}, obtained by adapting the gauge technique to this problem \cite{Delbourgo:1977jc}.

Little is known about $V_{\alpha\mu\rho}^{\rm T}$.  Attempts have been made in Abelian gauge theories to construct the analogous transverse part of the gauge-boson--fermion vertex using transverse Ward-Green-Takahashi identities \cite{He:2000we, Pennington:2005mw, Qin:2013mta}.  This approach has been extended to QCD \cite{He:2009sj, Chang:2010hb, Williams:2014iea, Aguilar:2014lha, Binosi:2016wcx, Aguilar:2016lbe, Bermudez:2017bpx, Cyrol:2017ewj}; and could, perhaps, be adapted to building a realistic model for $V_{\alpha\mu\rho}^{\rm T}$.   Meanwhile, however, as with all preceding continuum studies, we assume that $V_{\alpha\mu\rho}^{\rm T}$ does not contribute materially to the $0^{-+}$ Bethe-Salpeter kernel; or, at least, that its contribution may effectively be absorbed into the values of parameters used to define other aspects of any model input.  This assertion cannot now be validated, but that will change as further resources are devoted to analysing the three-gluon vertex \cite{Aguilar:2013vaa, Blum:2014gna, Eichmann:2014xya, Athenodorou:2016oyh, Duarte:2016ieu, Boucaud:2017obn, Corell:2018yil}.

Note that we have not discussed the massless poles which must appear in $V_{\alpha\mu\rho}$ in order to ensure emergence of a nonzero infrared mass-scale in QCD's gauge sector \cite{Aguilar:2015bud}.  This is because they couple longitudinally to all external lines; hence, owing to Eq.\,\eqref{BSTransverse}, are eliminated from the $0^{-+}$ channel.

With the three-gluon vertex thus defined, one may proceed to evaluate ${\mathpzc K}$ in Eq.\,\eqref{KKernel}, arriving at the following compact result:
\begin{align}
{\mathpzc K} &= T_1
\Big[ V_1 T_2 Q^2
+ 2 V_5 \left[(k\cdot P)(q\cdot P)-P^2(k\cdot q)\right]\Big]
\nonumber \\
&+2T_2
\left[ 2 V_3\, (k\cdot Q) - 2 V_2\, (q\cdot Q) +V_4\, (P\cdot Q)\right]
\,,
\label{K1s}
\end{align}
where
\begin{subequations}
\label{t12}
\begin{align}
T_1&= \frac{1}{Q^2}\left[P^2Q^2 -(P\cdot Q)^2  +4(k\cdot q)^2- 4k^2q^2\right],\\
T_2&=\frac{1}{Q^2}\left[k^2(q\cdot P)^2-k^2q^2P^2+ P^2(k\cdot q)^2\right.  \nonumber \\ &\left. \hspace{1.5cm}+ q^2(k\cdot P)^2-2(k\cdot P)(k\cdot q)(P\cdot q)\right];
\end{align}
\end{subequations}
\begin{subequations}
\label{comb3g}
\begin{align}
V_1&:=(X^-_3+X_6^-+X^-_9)\,\widetilde{X}^+_1 \nonumber\\
    & \qquad + (X_3^+ +X_6^++X_9^+)\,\widetilde{X}_1^- , \\
V_2&:=\widetilde{X}_4^-(\widetilde{X}_1^+ + \widetilde{X}_7^+) + \widetilde{X}_4^+(\widetilde{X}_1^{-} + \widetilde{X}_7^-) , \\
V_3&:= \widetilde{X}_7^+(\widetilde{X}_1^- + \widetilde{X}_4^-) + (\widetilde{X}_1^+ + \widetilde{X}_4^+)\widetilde{X}_7^- , \\
V_4&:= \widetilde{X}_1^{-}(\widetilde{X}_4^+ - \widetilde{X}_7^+)- \widetilde{X}_1^+(\widetilde{X}_4^- -\widetilde{X}_7^-) ,\\
V_5&:=\widetilde{X}_1^+\widetilde{X}_1^-\,;
\end{align}
\end{subequations}
and
\begin{subequations}
\label{combi}
\begin{align}
\widetilde{X}_1(q,r,p)&=X_1(q, r, p)-(q\cdot r)X_3(q, r, p) \,,\\
\widetilde{X}_4(q,r,p)&=X_4(q, r, p)-(p\cdot r)X_6(q, r, p)\,,\\
\widetilde{X}_7(q,r,p)&=X_7(q, r, p)-(p\cdot q)X_9(q, r, p) \,,
\end{align}
\end{subequations}
with the arguments of the functions $\widetilde{X}_i^{\pm}$ being obtained from Eq.\,\eqref{combi} via appropriate identification of $q$, $r$, $p$.

It is worth highlighting that ${\mathpzc K}$ in Eq.\,\eqref{KKernel} is symmetric under \mbox{$+ \leftrightarrow -$}, \emph{i.e}.\ invariant under the simultaneous operations $q\rightarrow -q$, $k\rightarrow -k$; consequently, $Q\rightarrow -Q$.  This follows because $\{V_i , i=1,2,3,5\}$ are symmetric under the exchange operations and $V_4$ is antisymmetric.

\smallskip

\noindent{\sf \textbf{3$\;$Completing the Bound-State Kernel}}.
To compute the mass and bound-state amplitude for the $0^{-+}$ glueball it is necessary to complete the kernel of Fig.\,\ref{fig:BSE} by specifying forms for: (\emph{i}) the propagator of the valence-gluon constituents, $D({\mathpzc l}^2)$; (\emph{ii}) the functions determining the contributing parts of the three-gluon vertex, $\{X_{1,3,4,6,7,9}\}$; and the exchange interaction that binds the system, \emph{i.e}.\ the ladder rung in Fig.\,\ref{fig:BSE}.  We now consider each in turn.

(\emph{i}) -- $D({\mathpzc l}^2)$.  Following the pioneering effort in Ref.\,\cite{Cornwall:1981zr}, much has been learnt about the nature of the dressed-gluon two-point function in Landau gauge as continuum and lattice studies of QCD's gauge sector have increased in sophistication and reliability.  Today it is known that $D({\mathpzc l}^2)$ is well described by its perturbative form on \mbox{${\mathpzc l}^2\gtrsim 1\,$GeV$^2$}.  On the other hand, this propagator saturates at infrared momenta \cite{Cornwall:1981zr, Boucaud:2006if, Aguilar:2008xm, Boucaud:2011ug, Binosi:2014aea, Aguilar:2015bud, Cyrol:2016tym, Binosi:2016wcx, Binosi:2016nme, Gao:2017uox, Cyrol:2017ewj, Rodriguez-Quintero:2018wma}:
\begin{equation}
\label{eqGluonMass}
D({\mathpzc l}^2\simeq 0) = 1/m_g^2,
\end{equation}
as a consequence of gluon self-interactions.  Hence, a large body of work can be summarised by stating that gluons, although acting as massless degrees-of-freedom on the perturbative domain, actually possess a running mass, whose value at infrared momenta is characterised by $m_g$.

Such behaviour is expressed in each one of the many available numerical solutions of the gauge-sector gap equations.  However, those solutions are typically restricted to ${\mathpzc l}^2>0$, whereas solving the Bethe-Salpeter equation in Fig.\,\ref{fig:BSE} requires that $D({\mathpzc l}^2)$ be sampled on a sizeable subdomain of the entire complex ${\mathpzc l}^2$-plane \cite{Maris:1997tm}.  Any continuation into the complex plane of the known solution $D({\mathpzc l}^2>0)$ will typically depend sensitively on the truncations used for the relevant gap equations and its use in solving the Bethe-Salpeter will demand a high degree of numerical complexity.  Both effects place hurdles in the path leading to development of physical insights.

We therefore elect to follow Ref.\,\cite{Xu:2018cor} and employ a simple algebraic \emph{Ansatz} for the gluon propagator, which qualitatively preserves its known features and extrapolates into the complex plane in a manner consistent with confinement:
\begin{equation}
D({\mathpzc l}^2) = \frac{1}{m_g^2}\frac{1 - {\rm e}^{-{\mathpzc l}^2/m_g^2}}{{\mathpzc l}^2/m_g^2}=: {\mathpzc E}({\mathpzc l}^2,m_g^2)\,.
\label{GluonProp}
\end{equation}

It is appropriate here to explain that we view confinement as being effected by marked, dynamically-driven changes in the analytic structure of coloured Schwinger functions.  The modifications ensure that such functions violate the axiom of reflection positivity and thereby entails elimination of the associated excitations from the Hilbert space of asymptotic states \cite{GJ81}.  This is a sufficient condition for confinement  \cite{Gribov:1977wm, Gribov:1999ui, Munczek:1983dx, Stingl:1985hx, Krein:1990sf, Burden:1991gd, Hawes:1993ef, Roberts:1994dr, Roberts:2007ji, Strauss:2012dg, Qin:2013ufa, Lowdon:2015fig, Lucha:2016vte, Binosi:2016xxu, Binosi:2019ecz}, which leads one to view parton fragmentation phenomena as the cleanest expression of confinement in QCD \cite{Aguilar:2019teb}.

\begin{figure}[t]
 	\includegraphics[clip, width=0.45\textwidth]{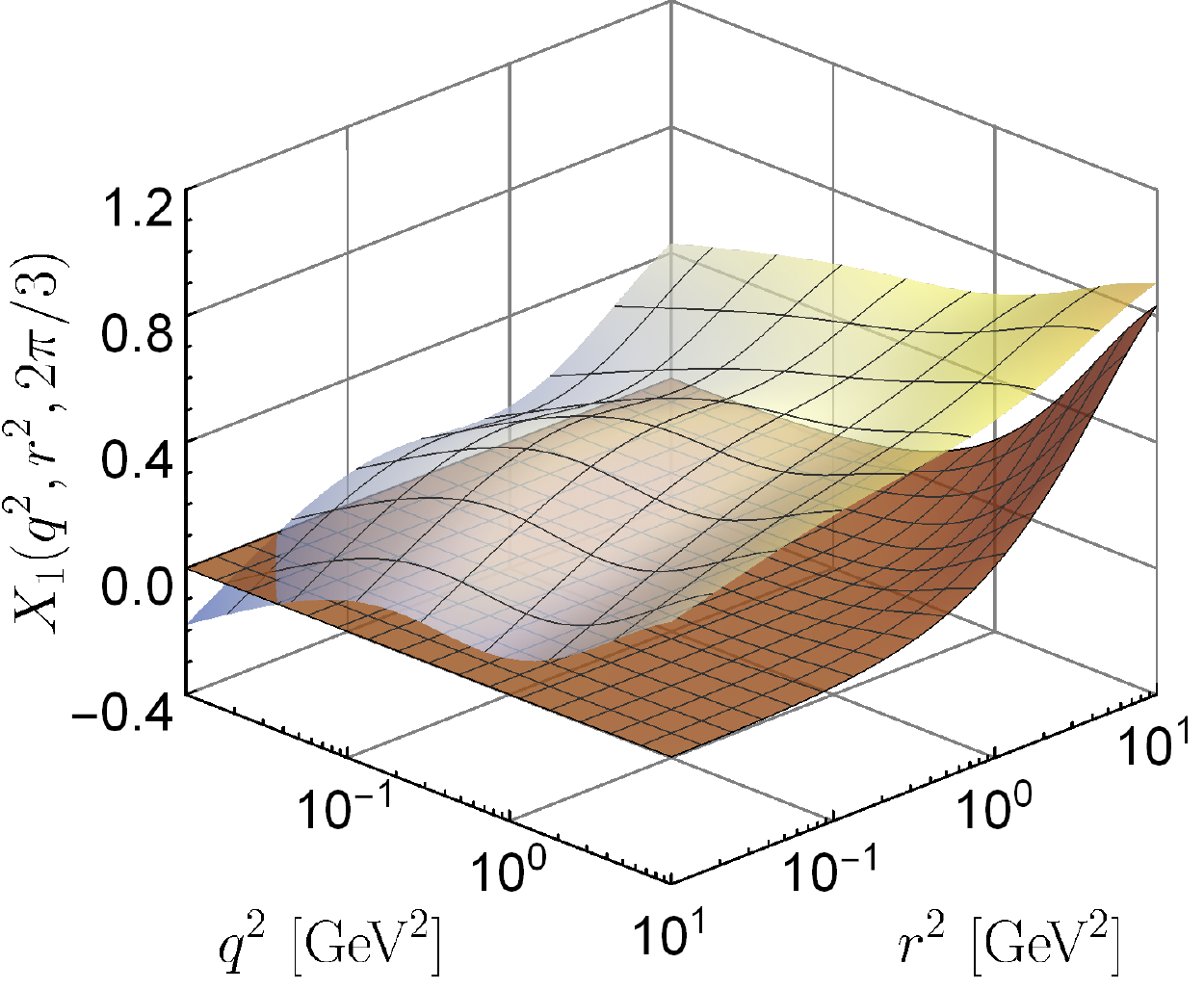}
\caption{\label{figf3g}
Vertex \emph{Ansatz} defined in connection with Eqs.\,\eqref{vertexAnsatz}, \eqref{f3g}, using $\omega_{3g}=1.9\,$GeV -- densely cross-hatched (brown) surface,  compared with a contemporary numerical solution for $X_1$ \cite{Aguilar:2019jsj} -- sparsely cross-hatched (blue-yellow) surface, plotted using a uniform angle of $2\pi/3$ between each of the four-vector arguments.}
\end{figure}

(\emph{ii}) -- $\{X_{1,3,4,6,7,9}\}$.  Above (see Sec.~2) we have sketched much of what is known about the three-gluon vertex.  It remains to add that at momentum scales above 1\,GeV$^2$, the vertex is once again well approximated by Eq.\,\eqref{Bare3Gluon}.  Here the dominant and striking nonperturbative effect is a marked suppression of the vertex at infrared momenta \cite{Aguilar:2013vaa, Blum:2014gna, Eichmann:2014xya, Athenodorou:2016oyh, Duarte:2016ieu, Boucaud:2017obn, Corell:2018yil}.  We implement these features by writing $X_{3,6,9}\equiv 0$,
\begin{subequations}
\label{vertexAnsatz}
\begin{align}
X_1^\pm & = X_4^\pm = X_7^\pm = R^\pm , \\
R^\pm & = {\mathpzc f}(k_\pm^2){\mathpzc f}(q_\pm^2){\mathpzc f}(Q^2)\,,
\end{align}
\end{subequations}
thereby preserving Bose symmetry of $V_{\alpha\mu\rho}$, with \cite{Xu:2018cor}
\begin{equation}
{\mathpzc f}(k^2) = 1 - \exp(-k^2/\omega_{3g}^2)\,,
\label{f3g}
\end{equation}
where the parameter $\omega_{3g}$ prescribes the domain over which suppression of the vertex is active.  Our approach replaces the full vertex by the bare vertex multiplied by the $R$ factor.  As illustrated by Fig.\,\ref{figf3g}, this algebraic \emph{Ansatz} qualitatively expresses the structure determined in modern numerical analyses of the three-gluon vertex.

(\emph{iii}) -- The exchange interaction in Fig.\,\ref{fig:BSE} describes a ladder-like approximation to gluon-gluon scattering.  As explained elsewhere \cite{Binosi:2016nme, Rodriguez-Quintero:2018wma}, by capitalising on a combination of the pinch technique \cite{Cornwall:1981zr, Cornwall:1989gv, Pilaftsis:1996fh, Binosi:2009qm} and background field method \cite{Abbott:1980hw, Abbott:1981ke}, one can define and compute a unique process-independent effective charge in QCD, \emph{i.e}.\ an interaction whose behaviour is identical in every scattering channel, gluon+gluon$\,\to\,$gluon+gluon, quark+quark$\,\to\,$quark+quark, \emph{etc}.  A phenomenologically efficacious representation of this interaction is explained in Refs.\,\cite{Qin:2011dd, Binosi:2014aea, Xu:2018cor}:
\begin{align}
\label{CalGQC}
{\mathpzc g^2}&(Q^2) D(Q^2)  \nonumber \\
& = \frac{8 \pi^2}{\omega^4} D \, {\rm e}^{-Q^2/\omega^2}
+ \frac{8 \pi^2 \gamma_m\, {\mathpzc E}(k^2,4\omega^2)}{\ln [ \tau + (1+Q^2/\Lambda_{\rm QCD}^2)^2]}\,,
\end{align}
with
$D\omega = (0.96\,{\rm GeV})^3$, $\omega = 0.5\,$GeV,
$\gamma_m = 12/25$, $\Lambda_{\rm QCD}=0.234\,$GeV, $\tau={\rm e}^2-1$.

We have now defined every element of the kernel depicted in Fig.\,\ref{fig:BSE}, enabling us to write the Bethe-Salpeter equation explicitly:
\begin{align}
\mathcal{F}(k;P) & = \frac{N_c}{2 h(k;P)}
\int \frac{d^4 q}{(2\pi)^4} \,{\mathpzc g^2}(Q^2) {\mathpzc K} \nonumber \\
& \quad \times D(q_+^2) D(q_-^2) D(Q^2)\, \mathcal{F}(q;P)\,,
\label{BSEfinal}
\end{align}
with $N_c=3$, \mbox{$h(k,P) :=[k^2P^2-(k \cdot P)^2]$} and
\begin{equation}
{\mathcal K} =\left\{2T_1\left[(k\cdot P)(q\cdot P)-P^2(k\cdot q)\right]
-16Q^2T_2\right\}R^{\,+}R^{\,-} .
\label{Kfinal}
\end{equation}
There is no realistic Bethe-Salpeter equation in hadron physics which takes a simpler form.

It is worth remarking here that if one writes $R^\pm \equiv 1$ in Eq.\,\eqref{Kfinal}, then Eq.\,\eqref{BSEfinal} simplifies to an equation with the character of that studied in Ref.\,\cite{Meyers:2012ka}.  Similarly, the $0^{-+}$ glueball bound-state equation analysed in Ref.\,\cite{Sanchis-Alepuz:2015hma} can also be recovered if one employs a different \emph{Ansatz} for the three-gluon vertex.
On the other hand, the analysis in Ref.\,\cite{Dudal:2010cd} does not follow a typical few-body approach to glueball structure, preferring instead to estimate glueball masses by studying the infrared behaviour of two-body composite operators built using a model for the gluon two-point function.  In this case, the resulting glueball masses are determined by the scales already contained in the gluon propagators because no dynamical information about gluon-gluon scattering is specified.

\smallskip

\noindent{\sf \textbf{4$\;$Solution for the $\mathbf 0^{-+}$ Glueball}}.
With every element in the kernel of Eq.\,\eqref{BSEfinal} given by an algebraic function, it is straightforward to solve for the bound-state mass and amplitude.  The symmetries explained in the ultimate paragraph of Sec.~2 ensure that the mass is real, despite the fact that the kernel is sampled in the complex plane.

Inspection reveals that the glueball mass increases as the mass of the valence-gluon, $m_g$, is increased: each valence-gluon propagator involves a $1/m_g^2$ factor that is dominant at infrared momenta; thus, any increase in $m_g^2$ must reduce the strength of couplings into the gluon-gluon correlation.  A similar effect is seen in the Bethe-Salpeter equation for any bound-state, even when a momentum-independent exchange-interaction is used \cite{Roberts:2011cf}; hence we do not discuss it further.  Instead, the value $m_g=0.6\,$GeV is taken from the hybrid-meson study in Ref.\,\cite{Xu:2018cor}: $\pm 10$\% variations in this value induce a response of commensurate size in the computed bound-state mass.

\begin{figure}[t]
 	\includegraphics[clip, width=0.44\textwidth]{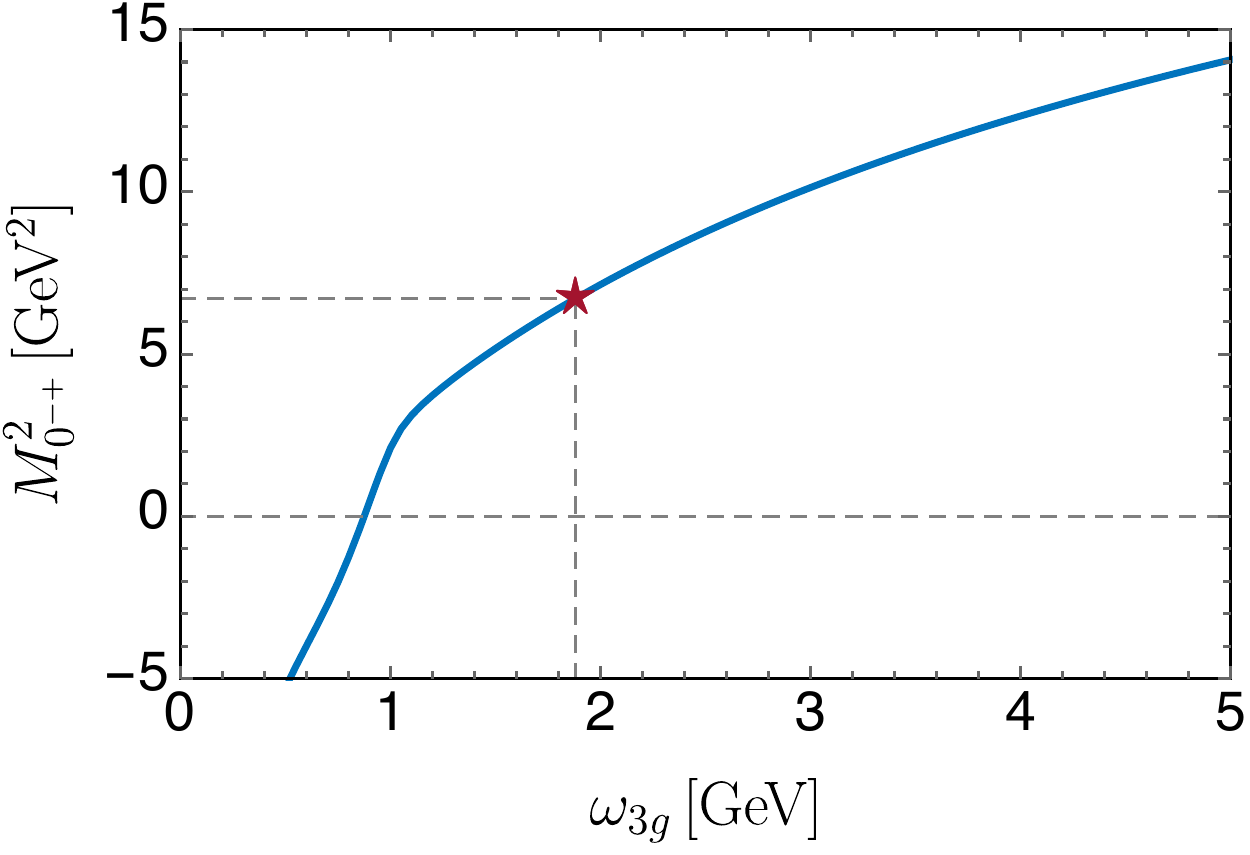}
\caption{\label{figMomega}
Mass-squared of the pseudoscalar glueball as a function of $\omega_{3g}$, whose value determines the size of the domain of infrared support for the kernel of the glueball Bethe-Salpeter equation: the size decreases as $\omega_{3g}$ increases.  The ``star'' marks the lQCD prediction from Ref.\,\cite{Chen:2005mg}: $M_{0^{-+}}=2.56(13)\,$GeV.}
\end{figure}

Our single parameter is $\omega_{3g}$, whose value prescribes the domain over which suppression of the three-gluon vertex is active in the Bethe-Salpeter kernel; and in Fig.\,\ref{figMomega} we display $M_{0^{-+}}^2(\omega_{3g})$, \emph{i.e}.\ the computed mass-squared of the $0^{-+}$ glueball as a function of $\omega_{3g}$.
Evidently, there is a critical value $\omega_{3g}^c = 0.87\,$GeV such that sensible solutions of Eq.\,\eqref{BSEfinal} are only obtained with $\omega_{3g}>\omega_{3g}^c$.
For $\omega_{3g}=\omega_{3g}^c$, there is so much attraction in the $0^{-+}$ channel that the lightest glueball appears as a composite zero mode; and the state becomes tachyonic for $\omega_{3g} < \omega_{3g}^c$.
This observation explains why Ref.\,\cite{Meyers:2012ka} found it necessary to employ a value of $\alpha(1\,{\rm GeV})$ that is just 1\% of the predicted result \cite{Binosi:2016nme, Rodriguez-Quintero:2018wma} in order to obtain a realistic value of $M_{0^{-+}}$.
On the other hand, $M_{0^{-+}}$ increases uniformly on $\omega_{3g}>\omega_{3g}^c$, with our result matching the lQCD prediction at $\omega_{3g} \approx 1.9\,$GeV.

We have repeated the analysis using the following propagator for the valence-gluon \cite{Gribov:1977wm, Stingl:1985hx}:
\begin{equation}
D_S({\mathpzc l}^2) = \frac{n_1^2}{d_1^4+({\mathpzc l}^2+d_2^2)^2}\,,
\label{PropStingl}
\end{equation}
$n_1=3.3$\,GeV, $d_1=0.99\,$GeV, $d_2=1.3\,$GeV, whose parameters were chosen to maximise the similarity between this function and that in Eq.\,\eqref{GluonProp} on the ray ${\mathpzc l}^2\geq 0$.  All results obtained using Eq.\,\eqref{PropStingl} are qualitatively and semiquantitatively equivalent to those already reported herein: in this case, the lQCD prediction for $M_{0^{-+}}$ is recovered using $\omega_{3g} \approx 1.7\,$GeV.

The behaviour of the solution trajectory in Fig.\,\ref{figMomega} can also explain the value of $M_{0^{-+}} \approx 4.5\,$GeV obtained in Ref.\,\cite{Sanchis-Alepuz:2015hma}.  Namely, the vertex \emph{Ansatz} employed therein likely provides too much infrared suppression in the Bethe-Salpeter kernel, \emph{viz}.\ is effectively associated with a large value of $\omega_{3g}$; a possibility implied in Ref.\,\cite{Sanchis-Alepuz:2015hma}.
%

\begin{figure}[t]
 	\includegraphics[clip, width=0.45\textwidth]{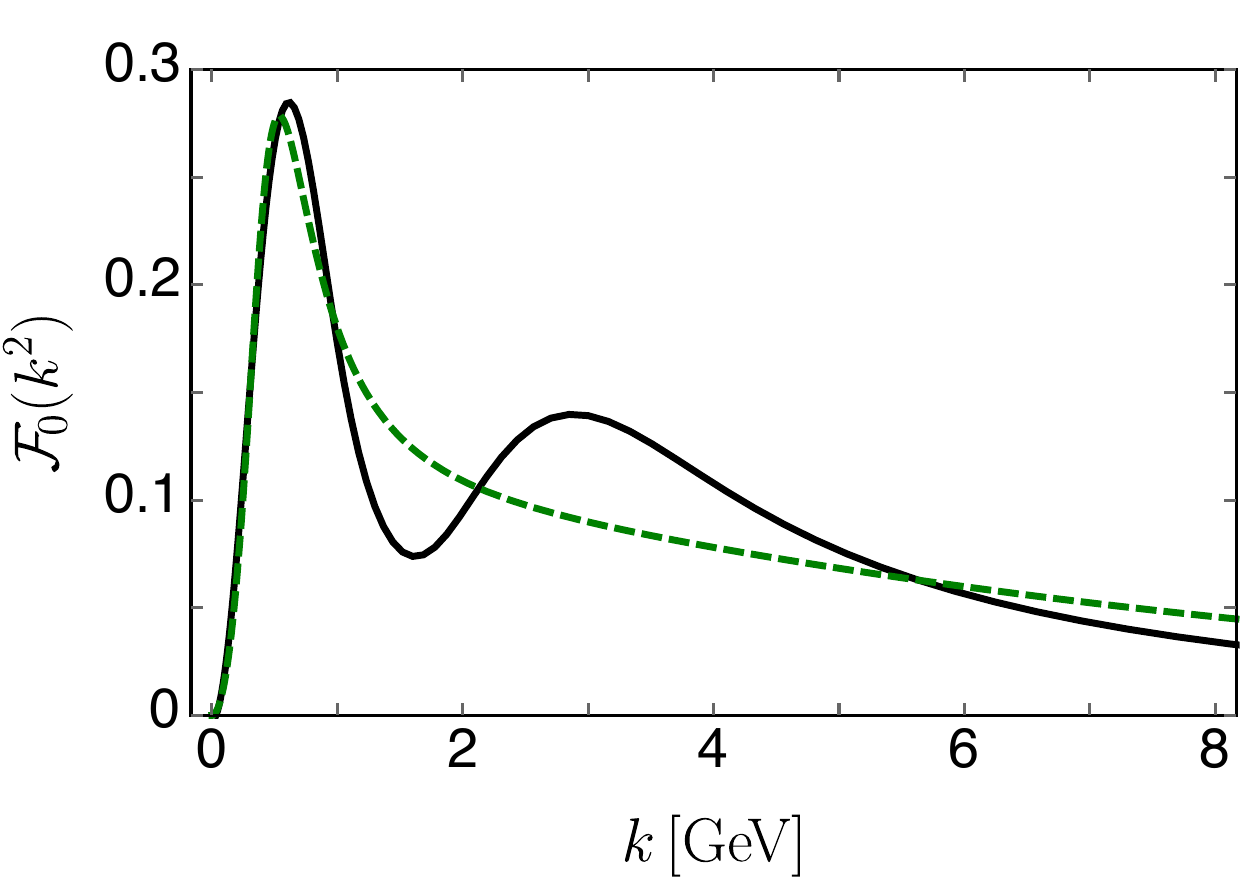}
\caption{\label{figBSA}
Zeroth Chebyshev moment of the $0^{-+}$ glueball Bethe-Salpeter amplitude, $\mathcal{F}(k;P)$, defined in Eq.\,\eqref{ZCM}.  Solid (black) curve -- directly computed result; and dashed (green) curve -- smoothened interpolation. }
\end{figure}

In Fig.\,\ref{figBSA} we plot the zeroth Chebyshev moment of the $0^{-+}$ glueball's Bethe-Salpeter amplitude:
\begin{equation}
\mathcal{F}_0(k^2) = \frac{2}{\pi}\int_{-1}^1dx\,\sqrt{1-x^2} \mathcal{F}(k;P)\,,
\label{ZCM}
\end{equation}
where $x=k\cdot P/\sqrt{k^2 P^2}$, calculated using Eqs.\,\eqref{GluonProp}\,--\,\eqref{Kfinal} with $\omega_{3g}=1.9\,$GeV.
$\mathcal{F}_0(k^2)$ vanishes in the neighbourhood $k^2\simeq 0$, grows to a peak and then asymptotically falls $\sim 1/k^2$, up to logarithmic corrections.

We emphasise that the local minimum and maximum at intermediate momenta are artefacts; also evident in the solution obtained using Eq.\,\eqref{PropStingl}.  They appear because we employed simple algebraic models for the gluon propagator, Eq.\,\eqref{GluonProp}, and three-gluon vertex, Eq.\,\eqref{vertexAnsatz}, \eqref{f3g}, when constructing the Bethe-Salpeter kernel instead of self-consistently determined Dyson-Schwinger equation solutions.  Consequently, mass-scales present in
Eqs.\,\eqref{GluonProp}\,--\,\eqref{CalGQC}
compete with each other instead of working together.  The dashed green curve is a least-squares smoothening fit to the numerical solution, which we expect to be a more realistic sketch of $\mathcal{F}_0(k^2)$.  This conjecture can be tested in subsequent studies that are built upon numerically determined kernel inputs.

The profile depicted in Fig.\,\ref{figBSA} is typical of the amplitude for a system with significant rest-frame orbital angular momentum between the dressed-valence constituents, as can be inferred by analogy with $P$-wave quark-antiquark mesons \cite{Li:2016mah}.
In the present case, owing to the three-gluon vertex, the kernel in Eq.\,\eqref{BSEfinal} has many numerator factors $\sim k\cdot P$, $q\cdot P$, \emph{etc}.  One must therefore anticipate a solution $\mathcal{F}(k^2,k\cdot P;P^2)$ with strong $k\cdot P$-dependence.  Furthermore, since a $0^{-+}$ state is even under $k\cdot P \to \mbox{$-k\cdot P$}$, then large relative angular momentum works to limit support in the neighbourhood $(k\cdot P)^2\simeq 0$. Consequently, the zeroth Chebyshev moment is suppressed on $k^2\simeq 0$.

It is natural to ask after the impact on our results of additional contributions to the gluon-gluon scattering kernel, \emph{i.e}.\ including additional terms on the right-hand-side of Fig.\,\ref{fig:BSE}.  In answer, we note that, in the pure-glue theory, any new term will involve at least two additional three-gluon vertices; hence, even greater suppression at infrared momenta.  Moreover, the support of such a contribution at ultraviolet momenta will be weaker than that of the leading term.  Consequently, whilst the results obtained from Eq.\,\eqref{BSEfinal} will receive minor quantitative modifications, they will be qualitatively unchanged.

\smallskip

\noindent{\sf \textbf{5$\;$Summary and Perspective}}.
%
Owing to the nature of gauge-sector dynamics, the Bethe-Salpeter equation describing gluon-gluon interactions in the $J^{PC}=0^{-+}$ glueball channel of pure-glue QCD takes a very simple form.  In fact, considering only the leading contribution to the associated kernel, it is arguably the simplest bound-state equation in hadron physics.

The kernel in the $0^{-+}$ channel is defined by a convolution involving the process-independent effective charge, dressed-gluon propagator, and three-gluon vertex: $V_{\mu\nu\rho}$.  The charge is immutable and the solution of the Bethe-Salpeter equation is largely insensitive to details of the dressed-gluon propagator, which in any event is well constrained by continuum and lattice studies.  In consequence, a calculation of the $0^{-+}$ glueball mass and bound-state amplitude serves as a fairly transparent window onto the character and form of the dressed three-gluon vertex.  This is a unique quality of the $0^{-+}$ glueball problem.  In contrast, for example, the role of the three-gluon vertex in the three-quark baryon problem is obscured by other effects, such as the formation of quark+quark correlations \cite{Segovia:2015ufa}; and although the three-gluon vertex does play an important role in determining the properties of hybrid mesons, complexities in this three valence-body problem work to mask its role \cite{Xu:2018cor}.

Given that there is much still to learn about $V_{\mu\nu\rho}$, the access provided by gluon-gluon scattering in the $0^{-+}$ channel elevates the importance of this problem: it becomes a valuable test-bed for \emph{Ans\"atze} and computed approximations for $V_{\mu\nu\rho}$.  Our study highlights this special role.  In particular, it focuses attention on the potential physical importance of a peculiar feature of contemporary results for $V_{\mu\nu\rho}$, \emph{viz}.\ the appearance of an infrared suppression (zero-crossing) in the scalar function that modulates the strength of the tensor structure associated with the bare three-gluon vertex after the impact of nonperturbative interactions is assessed  \cite{Aguilar:2013vaa, Blum:2014gna, Eichmann:2014xya, Athenodorou:2016oyh, Duarte:2016ieu, Boucaud:2017obn, Corell:2018yil}.

Modern theory indicates that the infrared suppression of $V_{\mu\nu\rho}$, whose effects we have studied herein, is an unavoidable consequence of \cite{Aguilar:2013vaa, Aguilar:2015bud, Athenodorou:2016oyh}: (\emph{i}) the emergence of a gluon running mass, which is large at infrared momenta; and (\emph{ii}) the nonperturbative absence of any such scale in the ghost sector, so that the ghost dressing function, $F(k^2)$, whilst finite, possesses a logarithmic branch point at $k^2=0$.  Our study suggests that this novel feature of $V_{\mu\nu\rho}$ has observable implications for the spectrum, decays and interactions of colour-singlet bound-states.  We showed this explicitly for the $0^{-+}$ glueball; and given the universality of interactions in QCD, what is true for one system should also be true in many others.

Having highlighted the potential importance and physical manifestations of infrared suppression in the three-gluon vertex, it is next worthwhile to confirm the results and conclusions presented herein using refinements of the \emph{Ans\"atze} used for the gluon propagator and three-gluon vertex.  For instance, using self-consistently determined solutions for these functions in the $0^{-+}$ channel.  Sensibly constrained, the extension of our study to include other glueball channels would also be valuable because it may lead to a unified, internally consistent and insightful understanding of the structure and emergence of glueballs; and perhaps, thereby, assist in solving the existence problem for quantum Yang-Mills theories.

\smallskip

%
\noindent{\sf \textbf{Acknowledgments}}.
We are grateful for insightful comments from D.~Binosi, L.~Chang, Z.-F.~Cui, C.\,S.~Fischer, F.~Gao and J.~Rodr{\'{\i}}guez-Quintero.
Work supported by:
Brazilian National Council for Scientific and Technological  Development (CNPq), under grant nos.\ 142226/2016-5, 305815/2015 and 464898/2014-5 (INCT-FNA);
Coordena\c{c}\~{a}o de Aperfei\c{c}oamento de Pessoal de N\'{\i}vel Superior - Brasil (CAPES) - Finance Code 001;
Federal Institute of Education, Science, and Technology of Piau{\'i} (PROAGRUPAR-IFRA 119/2018);
Generalitat Valenciana, under grant Prometeo II/2014/066;
Jiangsu Province \emph{Hundred Talents Plan for Professionals};
Nanjing University of Posts and Telecommunications Science Foundation, under grant no.~NY129032;
National Natural Science Foundation of China, under grant nos.~11847024 and 11905107;
S\~ao Paulo Research Foundation (FAPESP), under project nos.\ 2017/07595-0 and 2017/05685-2;
and
Spanish Ministry of Economy and Competitiveness (MINECO), under grant nos.\ FPA2017-84543-P and SEV-2014-0398. 


\end{document}